\documentclass[final,12pt,3p]{elsarticle}

\usepackage{amssymb}

\RequirePackage{fix-cm}

\usepackage{graphicx}

\begin{document}

\begin{frontmatter}

\title{A survey on the importance of visualization and social collaboration in academic digital libraries}

\author[label1]{Azam Majooni\corref{cor1}}
\address[label1]{Centre for Instructional Technology and Multimedia, Universiti Sains Malaysia, 11800, Minden, Malaysia }
\cortext[cor1]{Corresponding author}

\ead{azam.majooni@gmail.com}

\author[label1]{Mona Masood}

\author[label2]{Amir Akhavan}
\address[label2]{School of Computer Sciences, Universiti Sains Malaysia, 11800, Minden, Malaysia}

\begin{abstract}
From more than half a century ago indexing scientific articles has been studied intensively  to provide a more efficient data retrieval and to conserve researchers invaluable time. In the last two decades with the emergence of the World Wide Web and the rapid growth in the number of scientific documents online many academic databases and search engines were launched with almost similar structure in order to reduce the difficulty in finding, relating and sorting  of the existing scientific documents published online. The dramatic increase of the scientific documents in the last few years makes it necessary that the retrieved information by the search engines be analyzed and more organized and interpretable representation be displayed to the users. Information visualization is a great way for exploration of large and complex data sets, therefore it can be a natural candidate for the purpose of generating more comprehensible search results for the citation and academic databases. In this survey the usage pattern of the participants and their demands and ideas for the existence of other beneficial methods for literature review has been questioned and the results are quantitatively analyzed.

\end{abstract}


\end{frontmatter}


%
%
%
%
%
%
%




\section{Introduction}
\label{intro}

Garfield in a print in 1963 has discussed about the “Science Citation Index” and also has performed a very short review of the previous attempts done in order to provide a unified citation indexing system and believes that most of the unsuccessful attempts were due to financing, technology, inflexibility, and lack of imagination~\cite{Garfield1964}. His main purpose to design a Science Citation Index was to propose a new construction for scientific literature in order to create an information retrieval system. In the late nineties Eysenbach et al, (1998)  published a work on the management of medical information as one of the first ones to suggest a scientific database on the internet \cite{Eysenbach1998}. They also argued about the quality of information on the internet which makes the information sources on the internet unreliable especially for serious cases such as medical uses. After Eysenbach et al, 
with the evolution of the internet and the fast growth of the online scientific documents, different scientific databases such as the Web of Science (WoS), Scopus, Google scholar and many others were developed and comparison of academic databases, especially Google scholar, Scopus and Web of Science and PubMed have been conducted from various perspectives during the past few years. Winter et al. recently studied the expansion of the Google scholars versus Web of science with this concept that Google scholar is still in the process of updating and is developing rapidly~\cite{Winter2013}. Also according to Abrizah et al. there exists a lack of standard classification scheme  for a standard classification scheme for the journal indexing in both Web of science and Scopus and in general all the academic databases which makes ranking of the journals a difficult task \cite{Abrizah2012}.

 Therefore with emergence of scientific citation sources, academic digital libraries and rapid growth of information on the world wide web, the issue of measuring the quality and the importance of scientific papers became crucial factors and therefore infometrics attracted more attention~\cite{Bar-Ilan2007}. Until the early years of the 21st millennium, the only comprehensive tool for carrying out empirical research in infometrics was the ISI Citation Indexes, but slowly many other scientific citation indexing databases such as Google Scholar and Scopus joined this area. Majority of the researchers depend on the digital libraries and academic databases for their research activities, especially literature review~\cite{Borner2002}.\\

There are two major methods for conducting a literature review or finding a research trend. The first method needs more effort and is more time consuming, it depends on collecting, reading, summarizing of the available literature to find the trend of the research. The second method is based on the academic databases and in other words is the process of statistically analyzing publications from several points of view, such as institution, authors, journals and monitoring keywords using the citation indexing databases and academic databases~\cite{Wang2012a}.\\

There are two methods for conducting a literature review and finding the research trend. The first method needs more effort and is more time consuming, it depends on collecting, reading, summarizing of the available literature to find the trend of the research. The second method which is called `Bibliometrics' is based on the academic databases and in other words is the process of statistically analyzing publications from several points of view, such as institute, authors, journals, citations and monitoring keywords using the citation indexing databases and academic databases~\cite{Wang2012a}. Bibliometrics can be very useful in finding the trends of the research, significance of the study, emergence and development of the research field or and even a theory, effective variables in the improvement or shift in the flow of the research and so on~\cite{Raan2001}.

There were many efforts in achieving better ways of conducting fast literature review and understanding the connection between different studies several years before the emergence of the internet and rapid growth of the scientific papers. In the mid sixties Kessler suggested grouping scientific papers based on bibliographic coupling units ~\cite{Kessler1963} and Price ~\cite{DeSollaPrice1965} introduced the idea of the creating a network in order to outline the scientific papers. In this survey one of the goals is to find out importance of multiple variables such as visual contents, feedback and sorting techniques in the scientific digital libraries for the researchers. Aldo apart from literature review and research trend monitoring, academic databases can be very useful in measuring different factors such as predicting rate of migration of the researchers ~\cite{Moed2012}, evaluation of the scientific status of a country ~\cite{Gupta2009}, an institute ~\cite{Etxebarria2009}, departments and even researchers ~\cite{Torres-Salinas2009}.

One of the ways of representing the results of the Bibliometrics is by means of visual contents such as graphs (2D and 3D), diagrams, charts and infographics \cite{Schiebel2012}. One of the key parameters which has been in the scope of this survey is the effects and importance of visualization in the academic databases. Beside visualization, in this survey participants are asked about the importance and usefulness of the social collaboration, such as feedback, comments and scores like social networks, in academic databases and digital libraries. According to the hypothesis, existence of this facility can help the researchers in many different  ways also can help publishers to notify plagiarized contents quickly.

In a short overview, the organization of this paper is as follows. In section 2, aims and importance of the research is discussed, section 3 gives a brief review of the details of the survey conducted and in section 4, the results of the survey are demonstrated and analyzed and finally a short conclusion of the search results and discussion is made.

\section{Purpose of the survey}
\label{sec:1}

With the ever increasing number of scientific articles and emergence of various sources of information, quick and straightforward access to valid and reliable contents has become very important. One of the aims of this survey is to widen our knowledge about the user’s expectation from the academic databases. Therefore we try to collect information about user's behaviours and habits to identify their demands and potential favorite modifications in the interface and query results displayed in academic databases. These modifications can be useful for the majority of the users of the academic databases. Initially this survey tries to identify the most used academic databases and in the second phase tries to identify the expectations of the users from each of the databases by suggesting possible changes and facilities than can be provided by the databases. Particularly, evaluating view point of the higher education students and researchers about the issue of the visualization of the search results in academic databases has been emphasized in this survey. Analyzing the results of the survey try to predict usage pattern and popular trends among the researchers and specially higher education students, to find meaningful relations between different data-sets and examine willingness of the users of the academic databases towards different possible tweaks that the citation indexing databases can provide and discusses about them. 

\section{Methods}
\label{sec:2}

The design of the present study is based on a survey and data analysis scenario. The survey is designed to collect information about the usage pattern and estimation of the demand to specific factors such as visual contents, sorting options, feedback systems and popularity of scientific social networks.

\subsection{Survey Instrument}

The questionnaire consists of three major sections: first section asked for their academic and personal information, the second section was about the usage frequency and favorite databases and the third section consisted of questions about the existing facilities and possible updates in the academic databases. The survey was conducted both on the internet using online questionnaire designed and an identical paper based copy starting from early September 2013. In order to achieve a population of the participants related to the survey, the hard copy questionnaires has been distributed at academic events such as workshops and colloquiums inside ``Universiti Sains Malaysia'' and the soft copied were either emailed to postgraduate students or posted on academic groups in the social networks. The survey ended when the number of the participants reached to $105$ (it ran for almost $22$ days) and answers of $5$ of the participants were omitted as they were duplicates.\\

\subsection{Subjects and Recruitment}

In order to estimate the proper sample size with $95\%$ confidence level, statistical calculator of ``Creative Research Systems'' has been used \cite{surveysystem_com}.

For this survey with sample size of 105, confidence interval is almost equal to $8\%$ with the confidence level of $95\%$ and  without loss of generality, a large population size. Most of the questions of the survey where multiple choice with $4$ or more options giving enough suitable choices for all the participants.\\

\section{Results \& Discussion }
\label{sec:3}
The results of the survey are analyzed and the statistical comparison of some of the parameters which can be inferred from the questionnaire are discussed in this section.\\

\subsection{Distribution of the participants}

\begin{figure}[htbp]
	\centering
		\includegraphics[width=1\textwidth]{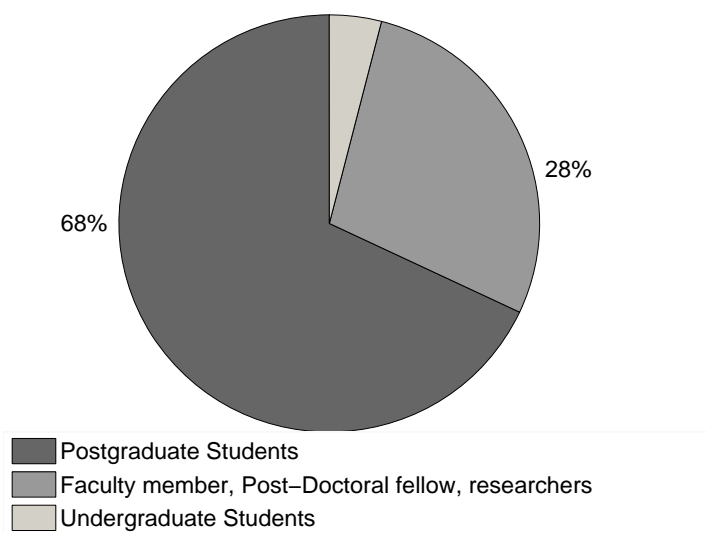}
	\caption{Academic status of the participants of the survey}
	\label{fig:academic_status}
\end{figure}

 From 100 participants, $68\%$  were post graduate students ($35\%$ PhD Candidates and $33\%$ MSc. Students), $28\%$ Faculty members, Post-Doctoral fellows and Researchers while $4\%$ were undergraduate students (Fig. \ref{fig:academic_status}). The distribution of the participants is denser on the postgraduate students which can help the research focus to be more on the postgraduate students.\\

As mentioned earlier, most of the participants of the study are postgraduate students. In this survey the age groups, academic field, frequency of access to the academic databases and many other parameters are questioned and the results are analyzed and discussed in this section.


\begin{table}
	\centering
			\caption{Age groups of the participants}
	\label{tab:AgeGroupsOfTheParticipants}
		\begin{tabular}{  l  c  }
Age &	Number of participants \\ \hline
$20$ to $25$ &	$16$ \\ \hline
$26$ to $30$ &	$33$ \\ \hline
$31$ to $35$ &	$25$ \\ \hline
$36$ to $40$ & 	$11$ \\ \hline
$41$ to $45$ &	$8$ \\ \hline
$46$ to $50$ &	$1$ \\ \hline
$51$ to $55$ &	$2$ \\ \hline
$56$ to $60$ &	$2$ \\ \hline 
$61$ to $65$ &	$1$ \\ \hline
$66$ to $70$ &	$1$\\ \hline

		\end{tabular}

\end{table}

Table ~\ref{tab:Disciplinestakingpartinthesurvey} demonstrates the age distribution of the participants and it can be seen that the majority of the participants are in the range of 20 to 35 years old, and 60\% of the participants are male versus 40\% females and the number of participants from Computer Sciences, I.T and Multimedia are noticeably higher than the other disciplines (See Table \ref{tab:AgeGroupsOfTheParticipants} and ~\ref{tab:Disciplinestakingpartinthesurvey}).


\begin{table}
	\centering
			\caption{Disciplines taking part in the survey}
	\label{tab:Disciplinestakingpartinthesurvey}
		\begin{tabular}{  l  c  }
Field &	Number of participants \\ \hline

Computer Science &	25 \\ \hline
Instructional Multimedia &	15 \\ \hline
Management	& 8 \\ \hline
Biology	& 6 \\ \hline
Medical	& 5 \\ \hline
Medicine &	4 \\ \hline
English Language &	4 \\ \hline
Education Technology &	4 \\ \hline
Physic &	3 \\ \hline
Chemistry &	3 \\ \hline
Engineering &	3 \\ \hline
Microbiology &	2 \\ \hline
Food Industry &	2 \\ \hline
Information Technology &	2 \\ \hline
Building Technology &	1 \\ \hline
Communication journalism &	1 \\ \hline
Earth Science &	1 \\ \hline
Economics &	1 \\ \hline
Environmental Science &	1 \\ \hline
Art &	1 \\ \hline
Gastroenterology &	1 \\ \hline
Structural Engineering &	1 \\ \hline
Geography &	1 \\ \hline
Health &	1 \\ \hline
Marine Science &	1 \\ \hline
Physics &	1 \\ \hline
Social Sciences &	1 \\ \hline
Finance &	1 \\ \hline
		\end{tabular}

\end{table}

\subsection{Multidisciplinary versus Subject and Publisher Specific academic databases}

\begin{figure}[htbp]
	\centering
		\includegraphics[width=1\textwidth]{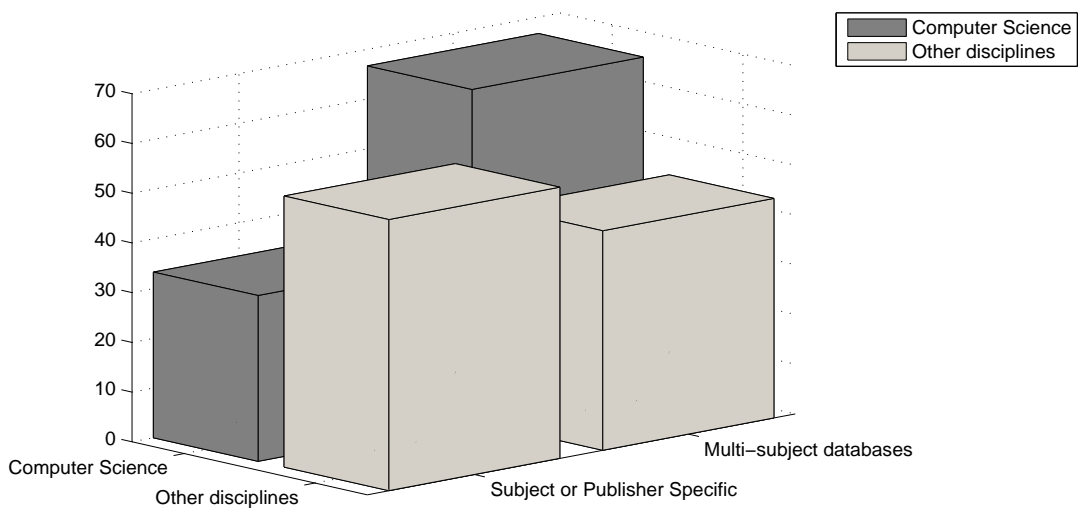}
			\caption{Multidisciplinary versus Subject and Publisher Specific academic databases}
	\label{fig:multi-cs-others}
\end{figure}

The results of the small scale survey demonstrates that there are considerable differences between the routine use of the academic databases from the discipline point of view. From all the academic fields taking part in this survey, participants from Computer Science had a slightly higher number of participants ($25\%$ of sample see Table \ref{tab:Disciplinestakingpartinthesurvey}). So that in this section we have tried to compare preference of the researchers from Computer Science with all the other fields to seek possible similarities. Figure ~\ref{fig:multi-cs-others} demonstrates the outcome of this comparison, it can be inferred that there exist diversity in the approach of the Computer Science researchers and other academic fields. This outcome, partially conflicts with the hypothesis that Multidisciplinary academic databases are the most proffered  and the subject specific and publisher specific databases, although with smaller coverage still have considerable amount of users between different fields. Also the statistical analysis of the survey results demonstrate there is no significant difference between the preference of Postgraduate students and Faculty members, Post-Doctoral fellows and researchers in the issue of the use of multidisciplinary versus subject and publisher databases.

\subsection{Most popular academic databases according to the survey results }


The popularity of an academic databases is a function with several parameters such as the number of articles, number of publishers, academic fields, subscription fees, facilities and ease of access and user friendliness of the interface. According to the statistics from the survey, academic databases which are more popular are listed in the table ~\ref{tab:PopularityOfTheAcademicDatabasesAndCitationIndexes}. Between the popular databases and scientific indexing websites, Sciencedirect and Springerlink has the most popularity with $87\%$ and $73\%$ of the participants respectively have been used at least once and $67\%$ and $47\%$ as regular academic search tools respectively.  Google scholar as scientific indexing tool is also very popular with $70\%$ of the participants having used it at least once and $40\%$ more regularly. The popularity of the other indexing tools and academic databases are mentioned in table \ref{tab:PopularityOfTheAcademicDatabasesAndCitationIndexes}. From the table, it can be inferred that, there are many users that have never even used some of the databases even once (or they have forgotten about it) for example $41\%$ of the participants have never used Scopus which is one of the largest and fastest bibliographic databases that can search above $20000$ journals~\cite{Hightower2010}, this can be due to subscription fees of the mentioned digital library.

\begin{table}[htbp]
	\centering
		\caption{Popularity of the Academic databases and Citation Indexes}
	\label{tab:PopularityOfTheAcademicDatabasesAndCitationIndexes}
\begin{tabular}{  l   c  c  }
Database name & Used at least once & Used regularly \\
\hline
	ScienceDirect & 87 & 67 \\ \hline
	Springerlink & 73 & 47 \\ \hline
	Google Scholar & 70 & 40 \\ \hline
	Scopus & 59 & 36 \\ \hline
	IEEE & 52 & 36 \\ \hline
	John Wiley \& Sons & 44 & 27 \\ \hline
	Proquest & 38 & 23 \\ \hline
	SAGE & 37 & 19 \\ \hline
	Ebscohost & 35 & 22 \\ \hline
	ACM & 34 & 18 \\ \hline
	PubMed & 30 & 25 \\ \hline
	Emerald & 27 & 16 \\ \hline
	Web of Science & 23 & 10 \\ \hline
	Mendeley & 14 & 2 \\ \hline
	citeSeerX & 13 & 4 \\ \hline
	BioMed & 12 & 2 \\ \hline
	Scirus & 12 & 2 \\ \hline
	Microsoft Academic & 10 & 3 \\ \hline
	PROLA & 6 & 2 \\ \hline
	WorldCat & 6 & 2 \\ \hline
	JSTOR & 6 & 3 \\ \hline
	arXiv & 5 & 1 \\ \hline
  IOP & 3 & 1 \\ \hline
	MathSciNet & 2 & 0 \\ \hline
\end{tabular}

\end{table}

\subsection{Frequency of use of academic databases by academicians}

\begin{figure}
	\centering
		\includegraphics[width=1\textwidth]{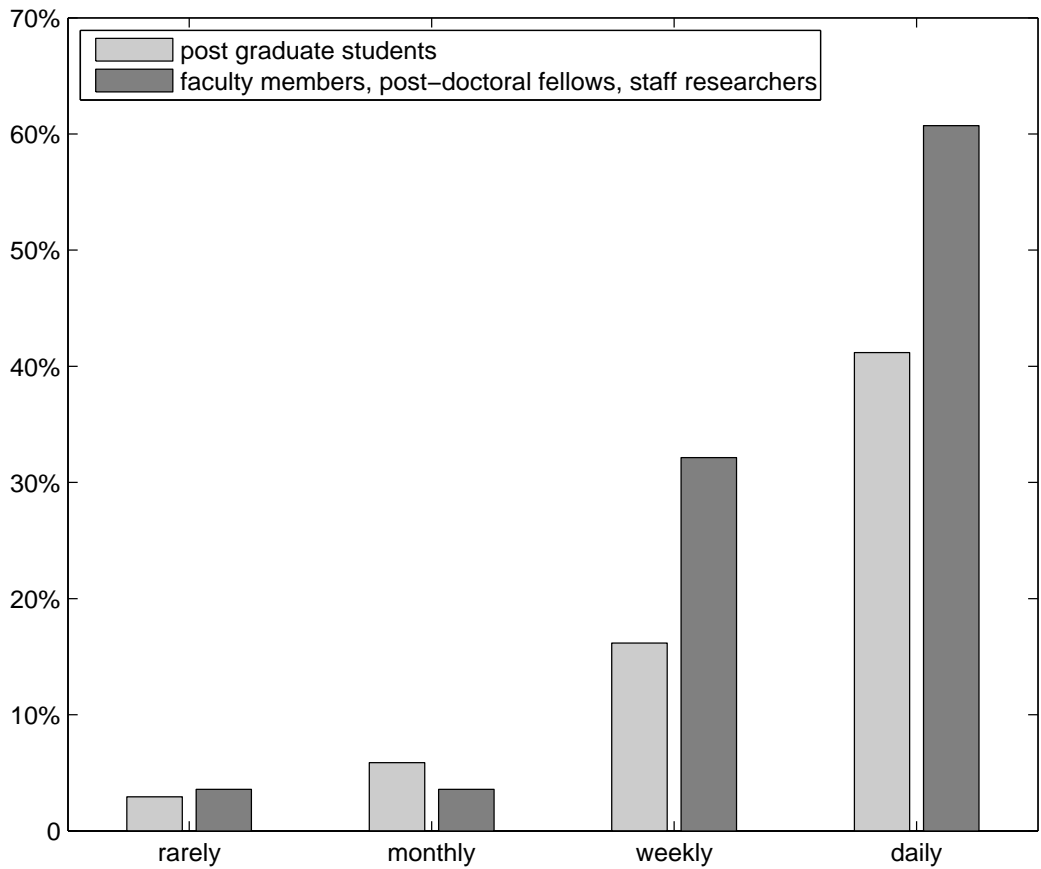}
			\caption{Frequency of use of academic databases by academicians}
	\label{fig:freq_phd_staff_all}
\end{figure}

In one of the sections of the questionnaire, the participants were asked about the frequency of the use of academic databases. Figure ~\ref{fig:freq_phd_staff_all} shows that most of the participants, either postgraduate, post-doctoral candidate, research staff, or faculty members use the academic databases quite regularly and most of the participants use them at least once a week. This result indicates the popularity of the academic databases between the postgraduate, post-doctoral candidate, research staff, or faculty members.

\subsection{Importance of graphical elements}

\begin{figure}
	\centering
		\includegraphics[width=1\textwidth]{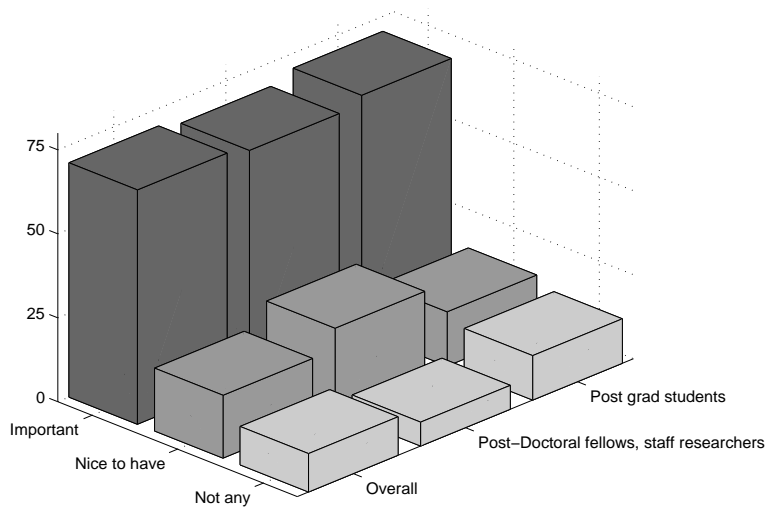}
			\caption{Academic status versus Demand to Graphics in the academic databases}
	\label{fig:importance_visual}
\end{figure}

Understanding and learning through visual contents is very much faster than the text and other methods of receiving information. The same concept can be true for the researchers trying to review the literature or understand connection between different concepts and disciplines. 
Figure \ref{fig:importance_visual} demonstrate that above $70\%$ of the participants pointed that the existence of visual contents and infographics in the databases are very important while less than $10\%$ found it not useful. The number $70\%$ with the margin of error of $7\%$for this specific survey, indicates existence of a demand for the visual contents inside academic databases.\\

\subsection{Importance of comments or a social communication system}

In this survey, participant's view point about the existence of a commenting system or a social communication system was elicited. The system allows the researchers to comment about an article or highlight and put notes inside them and accordingly their comments can be approved, scored, liked, replied or questioned. Such a system can be a revolutionary social network for the scientists and can be very important in the process of knowledge distribution and clarification of ambiguity in articles. 

\begin{figure}
	\centering
		\includegraphics[width=1\textwidth]{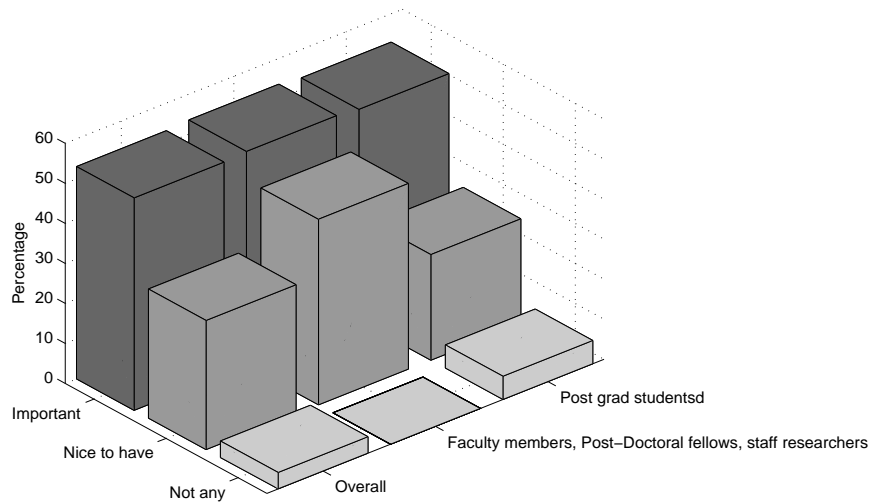}
		\caption{Academic status versus Demand to feedbacks, comments and ideas}
	\label{fig:importance_of_comments}
\end{figure}


\section{Importance of sorting capabilities in academic databases}

Scientific journals have a chief role in the scientific communication process. They have a regular pattern of publication and are typically focused on a specific field. Journal Articles are refereed, quality controlled and published based on a standard format \cite{Mayr1981}. Scholarly metrics such as impact factor, and h-index determine the ranking and in some cases importance the journal.\\

Morris in his paper~\cite{Morris2009} has modeled three scenarios which need mapping research and finding relevant information using bibliometrics. In the presented models the key questions are about: listing research topics, new emerging technologies in the fields, experts of the fields, centers of excellence, important journals that need to be monitored recommended reading list are asked and based on the answers a mapping system is suggested.\\

 In order to answer such questions, a researcher needs to preform a literature review over the topic and find precise results for this purpose~\cite{Mayr1981}. Searching for keywords between several databases can lead to a huge and complicated data-set and consequently big number of search results which can confuse the users. It is apparent that expectations of users searching the internet, and in this particular case, digital libraries are that the search engine list closest and most valuable articles therefore the results need to be ranked based on specific meters such as relevance of keywords and topics, impact factor of journal, h-index of the authors of the journal and number of citation. Therefore data retrieval systems use different methods to list the documents based on their relevance to the users query ~\cite{Mutschke2011,Mayr1981}.  There has been various methods of relevance ranking that has been applied in the search engines which are typically based on the frequency of the keywords versus documents frequency such as Google scholar besides that there exist several other methods for relevancy ranking, for academic databases when the number of search results are too many, such as Bradfordizing applied in the Web of Science.~\cite{Mutschke2011}. There has been several studies trying to investigate the users expectation while searching on the internet and particularly academic databases~\cite{Raan2001,Bar-Ilan2005,Scull1999,Bawden2006}.

In this survey, we have tried to point out the parameters which effectively influence behaviours, interests and demands of academic databases users. Sorting search results is one of the strategies in data clustering of small and large scale queries in order to narrow down the list to the most relevant results. As mentioned earlier, the results of the a search query on the academic databases are usually set to be sorted based on the relevance to the keyword searched by default~\cite{Bani-ahmad2008}. Scopus, Web of Science and Google scholar and other databases, give the possibility of sorting the results based on other methods such as date of publication, number of citation and name of the authors, affiliation besides the relevancy. In the survey, the participants are asked to mention the importance of each of the methods of sorting provided by academic databases. Later the results of the survey are analyzed. The results are presented in the Table ~\ref{tab:sorting} demonstrates that despite that the majority of the participants have found all the sorting options very important but sorting by authors is slightly more interesting for the researchers.

\begin{table}[htbp]
	\centering
		\caption{Comparison of importance of different types of sorting the queries in academic databases}
	\label{tab:sorting}

\begin{tabular}{  l  c   c  c  }
Sorting type & Important (\%) & Nice to have (\%) & Not important (\%) \\
\hline
date & 85 & 11 & 4 \\ 
\hline
 number of citation & 70 & 23 & 7 \\ 
\hline
title & 76 & 21 & 3 \\ 
\hline
impact factor & 66 & 30 & 4 \\ 
\hline
publication  & 62 & 30 & 8 \\ 
\hline 
affiliation  & 58 & 32 & 10 \\ 
\end{tabular}

\end{table}

Based on the availability and facilities, the participants are asked to choose score Google Scholar, Springerlink, Science direct, Web of Science Microsoft academic, Proquest and Pubmed based on their preference to choose. Overall results shows that Science direct and Google scholar and Springerlink had slightly higher scores which is in direct connection with the Sorting option, free access capabilities and distribution of participant's fields of research. Between the mentioned databases Web of Science has one of the best sorting and grouping tools and interfaces but it can be inferred that open access databases are usually more interesting for the researchers.

\section{Areas for Further Research}

It can be very useful to design a software or an application which can refine, visualize and make relation between search results of several academic databases and provide statistical conclusion about a specific searched keyword based on the dates, related papers, references, citations, authors and journals. Such a platform can be empowered with the ability of social networking basics so that the researchers can save more time with reading or seeing a social artifact provided by other users. The effect of visualization as mentioned earlier is significantly high and can give a big picture of huge pile of information and can be very informative in this case.
\section{Conclusion}
The analysis of the results of the survey gives an overall insight about the behaviours and habits of the users of the academic databases and also provide the information about their priorities, demands, requests for modifications in the interface and structure of the academic databases. Some of the databases, regardless of being very  rich in contents and being open access, attract limited number of users because of factors such as weaknesses in the user friendliness of the environment, lack of attention by the academic centers and universities, limited sorting and grouping facilities. In some cases there are even vivid problems and overlaps in the search results and exported citation data which consequently disappoints the users. This research gives the motivation for the future improvement in the academic databases, especially provides evidences about lack of visual contents and social collaboration in the available academic databases.
\section{Acknowledgments}
This work was supported by Centre for Instructional Technology and Multimedia, Universiti Sains Malaysia.\\
\bibliography{main} 
\bibliographystyle{elsarticle-num}

\end{document}